# INVESIGATION OF THE SERS SPECTRA OF COPPER PHTHALOCIANINE ADSORBED ON GALLIUM PHOSPHIDE SUBSTRATES

A.M. Polubotko* V.P. Chelibanov**

*A.F. Ioffe Physico-Technical Institute Russian Academy of Sciences, Politechnicheskaya 26, 194021 Saint Petersburg, Russia Tel: (812) 274-77-29, Fax: (812) 297-10-17  E-mail: alex.marina@mail.ioffe.ru

**State University of Information Technologies, Mechanics and Optics, Kronverkskii 49, 197101 Saint Petersburg, RUSSIA, E-mail: Chelibanov@gmail.com

## Abstract

The SERS spectra of the phthalocianine molecule, adsorbed on the gallium phosphide substrate are investigated. It is demonstrated that there appear strong lines, which are forbidden in usual Raman scattering. Analysis of the spectra indicates that these lines are associated with a strong quadrupole light-molecule interaction and also by a strong enhancement of the tangential components of the electric field on the surface. As it was demonstrated earlier, the last effect is characteristic for SERS on semiconductor and dielectric substrates, where there is the enhancement not only of the normal, but of the tangential components of the electric field on the surface.

Investigation of the SERS phenomenon on semiconductor and dielectric substrates is of a great interest both from the theoretical and experimental points of view. In [1] it was demonstrated that the reason of SERS in this case is the surface roughness. The enhancement arises in small regions of the surface with a very large curvature. As it was demonstrated in our work [1], the enhancement in SERS on semiconductor and dielectric substrates is weaker than on metals with the same value of the modulus of the dielectric constant. This result is associated with the fact that semiconductors and dielectrics are transparent in principle for the electromagnetic field in a wide range of frequencies in contrast with metals, which "expel" the field. Therefore the systems with the semiconductor and dielectric substrates possesses by a "smaller heterogeneity" of the medium, that results in a weaker enhancement of the field and its derivatives. However, in accordance with theoretical and experimental results [1], in contrast with the metal, there is the enhancement not only of the normal, but also of the tangential component of the electric field at the surface, that results in some peculiarities of the SERS spectra. We investigated the SERS spectra of the copper phthalocianine molecule (Figure 1), adsorbed on the gallium phosphide substrate, published in [2].

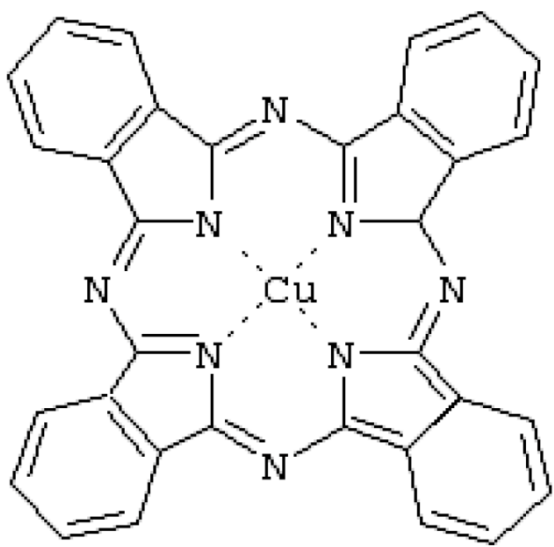

**Figure 1**. The Copper phthalocianine molecule

As it is well known copper phthallocianine belongs to the $D_{4h}$ symmetry group. The SERS spectra of this molecule, adsorbed on the particles of GaP with a mean sizes $d$ =106, 60, и 40 nm in the interval of the wavenumbers 500-1700 $см^{-1}$ are shown on the Figure 2.

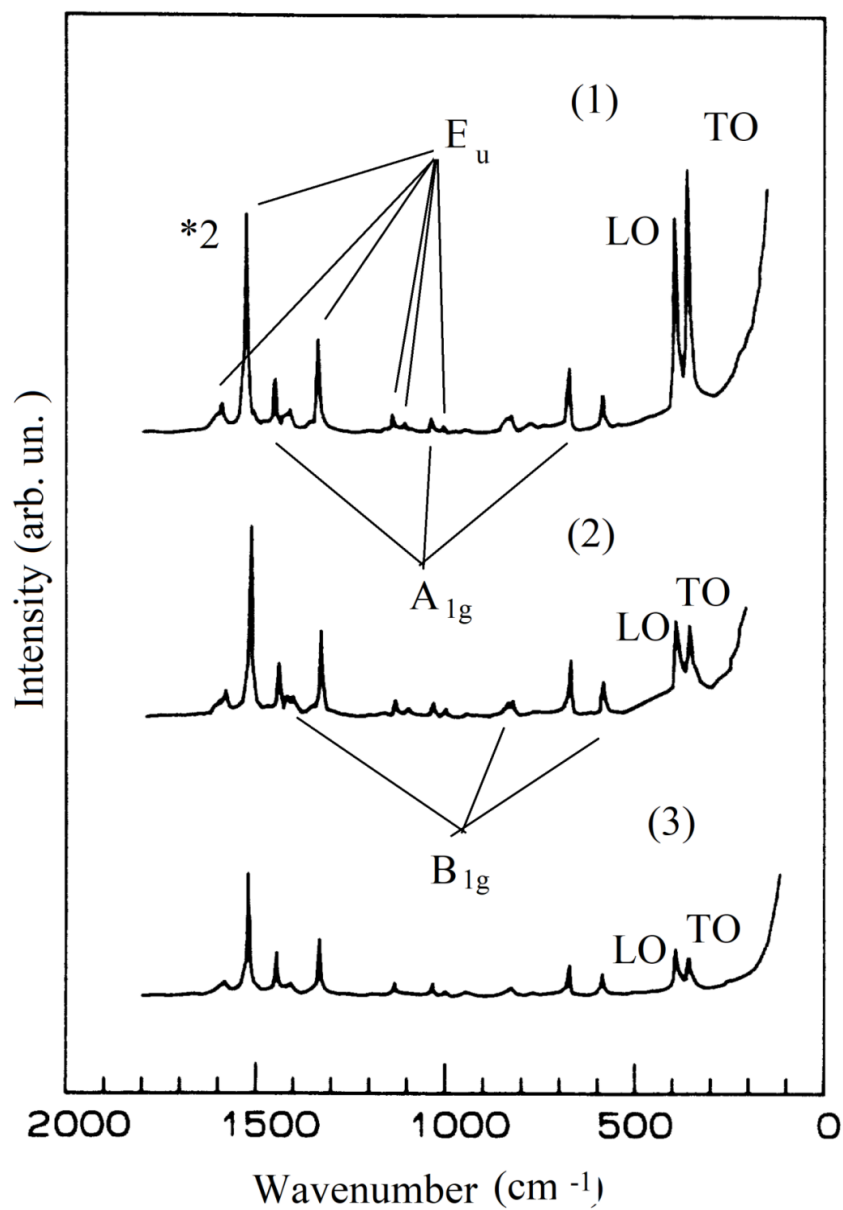

**Figure 2**. The SERS spectra of the phthalocianine molecule, adsorbed on the gallium phosphide substrate. (1) – the mean size of the particles 106 nm, the enhancement coefficient 700; (2) – the mean size of the particles 60 nm, the enhancement coefficient 300; (3) – the mean size of the particles 40 nm, the enhancement coefficient 200. LO and TO designate the lines, which belong to the longitudinal and transverse phonons of the substrate.

It is necessary to note that the spectra were taken for the wavelength of the incident radiation 514.5 nm. The enhancement coefficients were approximately 700, 300 and 200 respectively for the particles with the mean sizes $d =106$, 60, and 40 nm. In accordance with our ideas the phthalocianine molecule adsorbs parallel to the surface of nano particles. From Table 1 one can see, that there are sufficiently intense lines, caused by the vibrations with $A_{1g}$ and $B_{1g}$ irreducible representations and the wavenumbers 680, 1030, 1440 $cm^{-1}$ and 580, 1030, 1400 $cm^{-1}$ respectively, characteristic for the usual Raman scattering. However there appear strong forbidden lines with the irreducible representation $E_u$, caused by the vibrations transforming as the components of the dipole moments $(d_x, d_y)$ with the wavenumbers 1000, 1096, 1121, 1327, 1511 и 1580 $cm^{-1}$ and also a very weak line with the irreducible representation $A_{2u}$, caused by the vibration, transforming as the dipole moment $d_z$, which is perpendicular to the surface, with the wavenumber 950 $cm^{-1}$. Appearance of the forbidden lines with the irreducible representation $E_u$ is in conformity with our SERS theory on semiconductor and dielectric substrates [1], when there is the enhancement not only of the normal, but of the tangential components of the electric field. In accordance with our SERS theory [3], the scattering can occur via various dipole and quadrupole moments $d_x, d_y, d_z, Q_{xx}, Q_{yy}, Q_{zz}$, or via various linear combinations of the above quadrupole moments, transforming after the unit irreducible representation for the molecules, with sufficiently high symmetry. For the copper phthalocianine molecule these combinations are $Q_1 = Q_{xx} + Q_{yy}$ and $Q_2 = Q_{zz}$, which one names as main quadrupole moments $Q_{main}$. The scattering diagram for SERS is shown on the Figure 3. Here we do not repeat our reasoning, which one can find in our works and in the monograph [3] in particular. It is necessary only to indicate that the scattering contributions via dipole and main quadrupole moments $Q_1$ and $Q_2$, which we designate as $(d - Q_{main})$, define appearance of forbidden lines. Therefore their appearance, and also appearance of a weak line with the irreducible representation $A_{2u}$ indicate the appearance of a sufficiently strong quadrupole light-molecule interaction, which arises in the system with a semiconductor substrate in this case. One should note that in [4], we reported appearance of forbidden lines on the hydroquinone molecule, adsorbed on the $TiO_2$ substrate. However the forbidden lines with the irreducible representation $E_u$ have significantly higher intensity in our case, which is even more than the intensities of allowed lines with the irreducible representations $A_{1g}$ and $B_{1g}$, in contrast to the system with hydroquinone, where they are very weak. One of the reasons of this "anomaly" is the fact that the copper phthalocianine molecule is significantly larger than hydroquinone while the quadrupole interaction increases with the increase of the size of the molecules.

Table 1. Assignment of the lines of the copper phthalocianine molecule, adsorbed on the gallium phosphide substrate.

| Wavenumbers $cm^{-1}$ | Irreducible representations |
|---|---|
| 580 m | $B_{1g}$ |
| 680 m | $A_{1g}$ |
| 830 vw | $B_{1g}$ |
| 950 vw | $A_{2u}$ |
| 1000 vw | $E_u$ |
| 1030 vw | $A_{1g}$ |
| 1096 vw | $E_u$ |
| 1121 vw | $E_u$ |
| 1327 s | $E_u$ |
| 1400 vw | $B_{1g}$ |
| 1440 m | $A_{1g}$ |
| 1511 s | $E_u$ |
| 1580 w | $E_u$ |

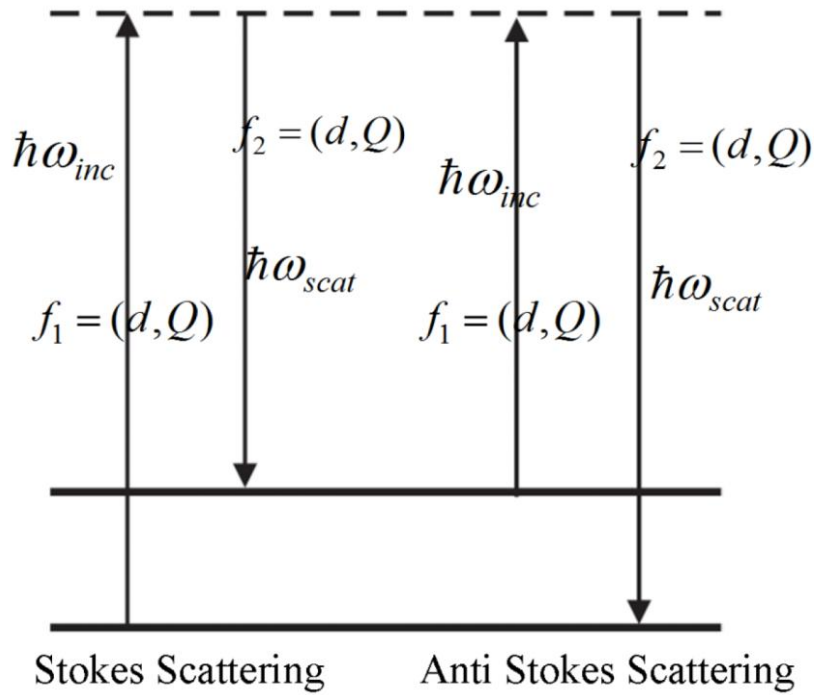

**Figure 3.** The scattering diagram for SERS. The scattering can occur via various combinations of the dipole and the main quadrupole moments $d$ and $Q$. Here $f_1$ and $f_2$ designate the dipole and main quadrupole moments, $\omega_{inc}$ and $\omega_{scat}$ are the frequencies of the incident and the scattered fields.

**References:**

1. Polubotko A.M., Chelibanov V.P. //Optics and spectroscopy, 2017, V. 122, № 6, P. 980.
2. Hayashi S., Koh R., Ichiyama Y., Yamamoto K.// Physical Review Letters, 1988, V. 60, No 11, P. 1085.
3. Polubotko A. M. The Dipole Quadrupole Theory of Surface Enhanced Raman Scattering, Nova Science Publishing Inc. 2009, New York.
4. Polubotko A.M. Chelibanov V.P.//Optics and spectroscopy, 2018, V. 124, № 1, P. 68.